\documentclass[fleqn,12pt,twoside]{article}
\usepackage{espcrc1}
\usepackage{graphicx}

\title{Photon and Electron induced Particle Production on Nuclei
\thanks{Work supported by DFG}}

\author{J. Lehr\address[theo]{Institut f\"ur Theoretische Physik, 
Universit\"at Giessen, Heinrich-Buff-Ring 16,\\ 35392 Giessen, Gemany}
and U. Mosel\addressmark[theo]}

\begin{document}

\maketitle

\section{Introduction}

Photon and electron induced production reactions on nuclei are of particular
interest. First of all, these probes are elementary and non-hadronic and 
allow therefore in principle to obtain information about the whole nuclear
volume.
Moreover, looking at different production channels, different fields of
physics are covered. To give a few examples,
meson photo- and electroproduction at lower energies
tells about the in-medium properties of nucleon resonances \cite{effe97,buu}.
Electron induced nucleon knockout reactions are closely linked to the
nucleon spectral function \cite{benhar,cda,corr} and also might give
experimental evidence of color transparency (see e.g. \cite{miller}).
In this text we want to give an overview over such reactions that can be 
described in our semiclassical BUU transport model.

\section{The Model}

\subsection{Description of Photon and Electron induced Processes}

In the usual one-photon exchange approximation electron induced reactions can
be treated theoretically in essentially the same way as photon induced
reactions. We assume that the photon (real or virtual) is absorbed
by a single nucleon inside the nucleus. This reaction is described by the
elementary cross sections, which in both cases actually only differ from
each other by an extra flux factor describing the electron vertex.
The particle content of the final states of this elementary reaction
varies with increasing photon energy from nucleon resonance excitation
and single/double pion production at lower photon energies, vector meson and 
strangeness production at larger energies. The relative importance of these
channels is given by the ratios of the corresponding cross sections.
At high energies, where more and more multiparticle channels open which have 
not been measured experimentally, we use the string fragmentation
model FRITIOF \cite{andersson}.

It is clear that these particles -- being produced somewhere inside the nucleus
-- are subject to final state interactions (FSI) due to the nuclear surrounding
and therefore the reliability of calculations of particle production cross
sections on nuclei depends crucially on how the FSI are treated.
The BUU model includes besides particle scattering and 
absorption also 'side-feeding' effects, contrary to other models 
(e.g. Glauber models).

The cross section for production reactions on nuclei in our model is given
by folding the cross sections of the elementary reaction on the nucleon
and the multiplicity of the particles under consideration reaching the
detector and an incoherent summation of the contributions from all nucleons
in the nucleus \cite{buu}.

\subsection{The BUU Model}

The model is based upon the BUU\footnote{BUU=Boltzmann, Uehling, Uhlenbeck}
equation, which for a system of nucleons is given by
\begin{equation}
  {\partial\over\partial t}f(\vec r,\vec p,t)+{\partial H\over \partial\vec p}
\vec\nabla_r f(\vec r,\vec p,t)-
{\partial H\over\partial\vec r}\vec\nabla_p f(\vec r,\vec p,t)=I_{coll}[f].
\end{equation}
Here $f$ and $H$ denote the nucleon phase space density and the mean-field
Hamilton function of the system, respectively. The left-hand side set to zero
yields the Vlasov equation, which gives a sufficient description of the
space-time evolution of a system of non-interacting particles. For a proper
consideration of interactions amongst the particles on the right-hand side 
a so-called collision integral is introduced.
Besides the nucleon our model contains a lot of other particle species, the
most important being nucleon resonances ($P_{33}(1232)$, $P_{11}(1440)$,
$D_{13}(1520)$, $S_{11}(1535)$...) and mesons ($\pi$, $\eta$, $\rho$, $K$,...)
(for a review of the model see \cite{buu}). 
For each of these species a BUU equation is introduced, which are coupled to
each other via the collisional integrals due to interactions between each 
other.
The usual method to solve the resulting set of coupled integral-differential
equations is to make a so-called test partice ansatz for the phase space
density $f$
\begin{equation}
  f(\vec r,\vec p)={1\over N}\sum_i\delta(\vec p-\vec p_i)\delta(\vec r-
\vec r_i). 
\end{equation}
For the propagation of the test particles this ansatz gives the classical
equations of motion
\begin{equation}
  \dot{\vec r_i}=\vec\nabla_p H,\ \dot{\vec p_i}=-\vec\nabla_r H.
\end{equation}

\section{Results}

\subsection{Meson Production}

Pion and eta production in the nucleon resonance region was calculated and 
discussed in detail in \cite{effe97,buu}. The applicability of the model 
extends
also to larger photon energies. However, for $E_\gamma > 1$ GeV the so-called
shadowing effect has to be taken into account. Since this is a coherence 
effect, it cannot be simulated in a semi-classical transport model.
In \cite{effe_gev} a procedure was developed to combine the description of 
these initial state interactions in the Glauber framework with the usual BUU 
model. Also the calculation of meson photoproduction at GeV energies can be 
found there.

\begin{figure}
\begin{center}
\vspace{-0.5cm}
    \includegraphics[width=12cm]{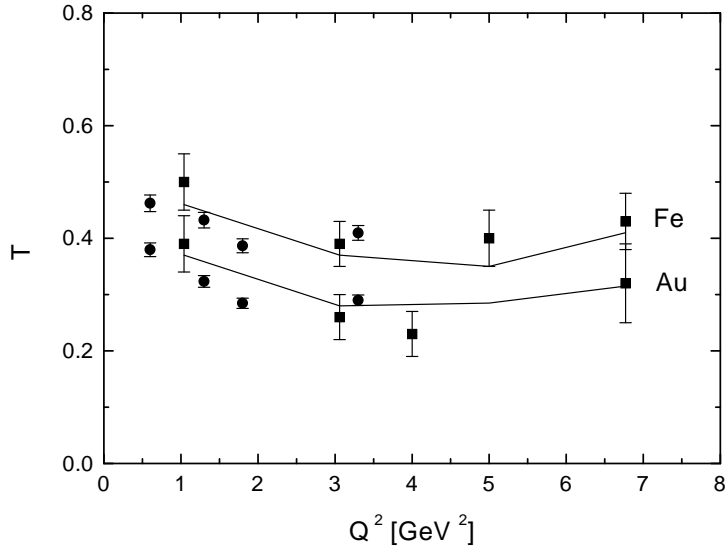}
\vspace{-1.5cm}
\end{center}
    \caption{$T$ as a function of $Q^2$.
The data are from \protect\cite{oneill} (solid) and 
\protect\cite{jlab} (open).}
\label{fig:el_knock}
\vspace{-0.5cm}
\end{figure}

\subsection{Electron induced Nucleon Knockout}

The electron induced nucleon knockout from nuclei (for a review see e.g.
\cite{frullani}) is of special interest
since it might give experimental evidence for color 
transparency.
The data from SLAC \cite{oneill} and TJNAF \cite{jlab} show 
the so-called transparency ratio which is given by the ratio of the number of 
protons measured close to the direction of the exchanged virtual photon and the
expected number of protons in the absence of FSI. Our model is capable to
provide both numbers consistently without further assumptions concerning the 
latter number. The transparency ratio was also calculated by other models in 
the past (see e.g. \cite{miller,golubeva} and references therein).

In Fig. \ref{fig:el_knock} we show our results for $T$ as a function of the
momentum transfer $Q^2$. For both nuclei Fe and Au we obtain very good
agreement with the data. 
Future work will also cover the region of larger $Q^2$ than available in the
experiments of \cite{oneill,jlab}, where an onset
of color transparency is expected and consequently an enhancement of $T$ 
towards one. Moreover, the influence of the nucleon spectral function on $T$
will be investigated.

\subsection{Nucleon Spectral Function and Subthreshold Particle Production}

Nucleons in nuclei may be off their mass shell in the 
nuclear groundstate due to short-range correlations (see e.g. 
\cite{benhar,cda,corr}). Calculations of the nucleon spectral functions
show that these groundstate correlations manifest themselves e.g. in the
momentum distribution. 
The Thomas-Fermi momentum distribution (usually used
by transport models to initialize the nucleon phase space density in
momentum space) is a step function $\Theta(p_F-p)$.
The effect of the short-range correlations is that some strength ($\sim$ 10\%)
is moved from momenta below $p_F$ to momenta above $p_F$.

\begin{figure}
\begin{center}
\vspace{-0.5cm}
\includegraphics[width=12cm]{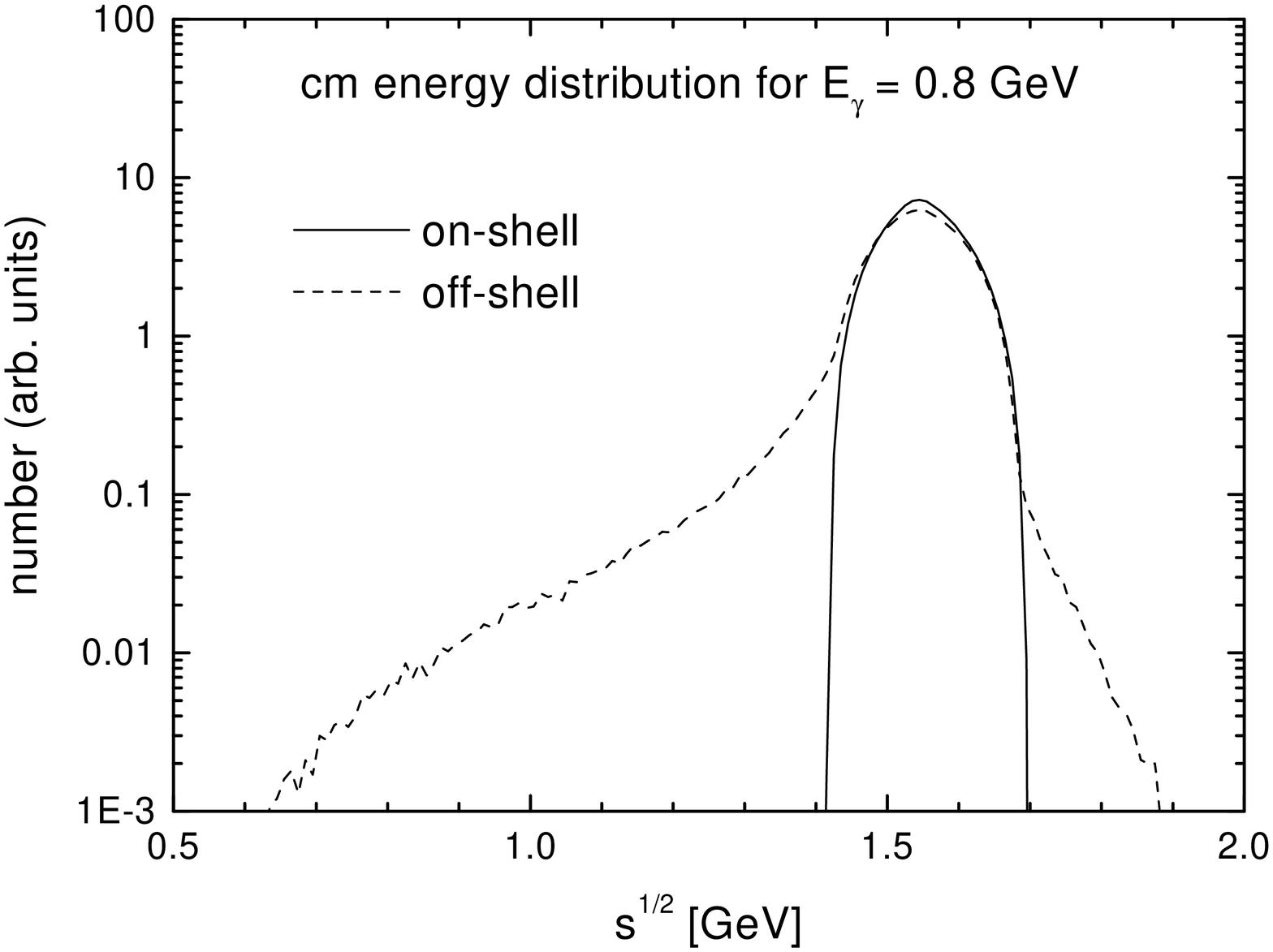}
\vspace{-1.5cm}
\end{center}
    \caption{CM energy distribution of $\gamma N$ pairs in $^{40}\textrm{Ca}$
for $E_\gamma=0.8$ GeV.}
\label{fig:srts}
\vspace{-0.5cm}
\end{figure}

The presence of short-range correlations has an influence on the particle
production close to threshold. This can be seen in Fig. \ref{fig:srts}, where
the center-of-mass energy distribution for the $\gamma N$ system involving
all nucleons in Calcium is shown for $E_\gamma=0.8$ GeV. The solid
curve shows the distribution caused by usual Fermi motion involving on-shell
nucleons. The dashed distribution, involving the full nucleon spectral 
function, is more smeared out and exhibits contributions for $\sqrt s$ 
\emph{larger} than the maximal value of $\sqrt s$ expected in the Thomas-Fermi
model. This is the region responsible for subthreshold particle production
effects.
In order to perform such calculations, our model has to be modified, because 
the nucleon phase space density has to be initialized differently than usual
and the off-shellness of the nucleons evolves during the reaction.
The latter fact is reflected in a modification of the equations of motion
for the test particles. A theoretical discussion of these issues and a 
derivation of the modified equations of motion can be found in \cite{leupold}.
A numerical implementation of off-shell nucleons in our transport model
for the description of heavy ion reactions was developed
in \cite{effe_off} and will also used as a basis for future work on the
description of subthreshold photoproduction.

\section{Summary and Outlook}

We have presented several examples for the calculation of inclusive particle 
production on nuclei within our semi-classical BUU transport model.
Future work will cover the extension of our calculations of $(e,e^\prime p)$
to larger $Q^2$ and photon induced reactions involving the nucleon spectral
function.

\end{document}